\documentclass[conference]{IEEEtran}
\IEEEoverridecommandlockouts
\usepackage{url}
\usepackage{cite}
\usepackage{amsmath,amssymb,amsfonts}
\usepackage{algorithmic}
\usepackage{graphicx}
\usepackage{textcomp}
\usepackage{xcolor}
\usepackage{booktabs}
\usepackage{threeparttable}
\usepackage{multirow}
\usepackage{marvosym}
\usepackage{colortbl}
\definecolor{deepgreen}{rgb}{0,0.7,0}
\def\BibTeX{{\rm B\kern-.05em{\sc i\kern-.025em b}\kern-.08em
		T\kern-.1667em\lower.7ex\hbox{E}\kern-.125emX}}

\begin{document}
\title{\hspace{0.41em}In-Loop Filtering via Trained Look-Up Tables \vspace{-0.3em}}
\author{
Zhuoyuan Li,
Jiacheng Li,
Yao Li,
Li Li,
Dong Liu\textsuperscript{\Letter},
and Feng Wu\\[0.2em]

\textit{University of Science and Technology of China (USTC),}

\textit{Hefei, Anhui 230027, China}\\[0.1em]

\{zhuoyuanli, jclee, mrliyao\}@mail.ustc.edu.cn, \{lil1, dongeliu, fengwu\}@ustc.edu.cn \vspace{-1em}

\thanks{
	\Letter: Corresponding author.}
}
\maketitle

\begin{abstract}
In-loop filtering (ILF) is a key technology for removing the artifacts in image/video coding standards. Recently, neural network-based in-loop filtering methods achieve remarkable coding gains beyond the capability of advanced video coding standards, which becomes a powerful coding tool candidate for future video coding standards. However, the utilization of deep neural networks (DNN) brings heavy time and computational complexity, and high demands of high-performance hardware, which is challenging to apply to the general uses of coding scene. To address this limitation, inspired by explorations in image restoration, we propose an efficient and practical in-loop filtering scheme by adopting the \textit{Look-up Table} (LUT). We train the DNN of in-loop filtering within a fixed filtering reference range, and cache the output values of the DNN into a LUT via traversing all possible inputs. At testing time in the coding process, the filtered pixel is generated by locating input pixels (to-be-filtered pixel with reference pixels) and interpolating cached filtered pixel values. To further enable the large filtering reference range with \textit{the limited storage cost of LUT}, we introduce the enhanced indexing mechanism in the filtering process, and clipping/finetuning mechanism in the training. The proposed method is implemented into the Versatile Video Coding (VVC) reference software, VTM-11.0. Experimental results show that the \textit{ultrafast}, \textit{very fast}, and \textit{fast} mode of the proposed method achieves on average 0.13\%/0.34\%/0.51\%, and 0.10\%/0.27\%/0.39\% BD-rate reduction, under the all intra (AI) and random access (RA) configurations. Especially, our method has friendly time and computational complexity, only 101\%/102\%$\sim$104\%/108\% time increase with 0.13$\sim$0.93 kMACs/pixel, and only 164$\sim$1148 KB storage cost for a single model. \textit{Our solution may shed light on the journey of practical neural network-based coding tool evolution.} 
\end{abstract}

\vspace{-0.7em}
\IEEEpeerreviewmaketitle

\vspace{-0.2em}
\section{Introduction}
\vspace{-0.1em}
In-loop filtering (ILF) has been widely adopted in modern video coding standards, including  H.266/VVC\cite{bross2021overview}, AV2\cite{AOM}. To promote the reconstruction quality of decoded frame, various complementary filters make a major contribution to these standards and play a key role in hybrid video coding framework, such as deblocking filter (DBF), sample adaptive offset (SAO), adaptive loop filtering (ALF)\cite{karczewicz2021vvc}. 

Recently, deep neural network-based (DNN) coding tools (e.g. intra prediction, ILF, etc.) have been rapidly developed\cite{liu2020deep, dai2017convolutional, dai2018cnn, li2023designs, li2021convolutional, li2021neural, li2017convolutional, li2020ahg11, EE1-1.6, EE1-1.7, zhao2023towards, ma2023overview, li2023idam, ding2023neural, kathariya2023joint}, and made good progress in some standardization activities, such as neural network-based video coding (NNVC)\cite{li2023designs}. The DNN-based tools sufficiently take advantage of data-driven capabilities to better fit the prediction or reconstruction goals. Although these deep tools have made impressive performance, they bring heavy time and computational complexity that makes them difficult to use in practice without high-performance hardware, and this is one of the major obstacles for practical deep tools.

To address this limitation, we propose an efficient and practical in-loop filtering scheme by adopting the \textit{Look-up Table} (LUT), which is inspired by explorations in image/video recovery tasks\cite{jo2021practical, li2022mulut, liu2023reconstructed, li2024toward, yin2024online}. The basic idea of the proposed scheme is to adopt the look-up operation (direct addressing) of LUT to replace the inference process of DNN in coding process, which is also friendly for embedded systems to accelerate computation with far fewer floating-point operations. To achieve this goal, we establish a LUT-based in-loop filtering framework (termed \textit{\textbf{LUT-ILF}}), and introduce a series of LUT-related modules to strengthen its efficiency, including the enhancement of filtering reference range with the limited LUT size (\textit{progressive indexing and reference indexing}, Section III), the optimization of LUT size with limited memory cost (\textit{clipping/finetuning}, Section II), the selection of reference pixels (\textit{learnable weighting}, Section III). 
Compared to the low/high complexity operation point setting (LOP/HOP) of NNVC-ILF\cite{AF0014, AF0041, AF0043}, our ultrafast mode (\textit{\textbf{LUT-ILF-U}}, reference range: 5$\times$5, 0.13 kMACs/pixel, 164 KB), very fast mode (\textit{\textbf{LUT-ILF-V}}, reference range: 9$\times$9, 0.40 kMACs/pixel, 492 KB) and fast mode (\textit{\textbf{LUT-ILF-F}}, reference range: 13$\times$13, 0.93 kMACs/pixel, 1148 KB) provide a series of new trade-off points that show lower time and computational complexity and good performance beyond VVC.

\begin{figure*}
	\centering
	\vspace{-1.9em}
	\hspace{-12mm}\includegraphics[width=182mm]{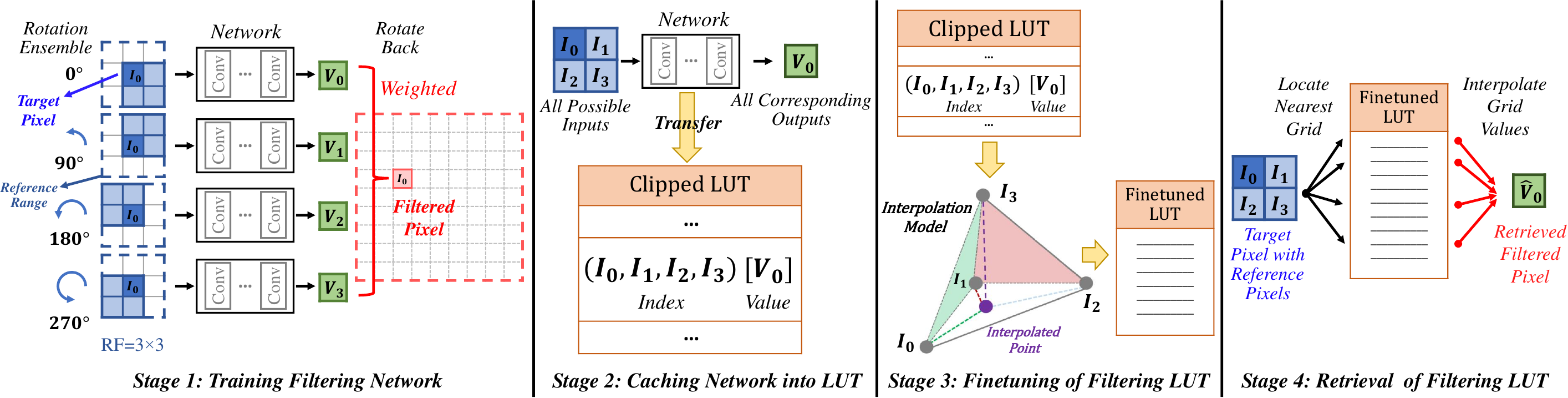}\hspace{-12mm}
	\vspace{-0.8em}
	\caption{Illustration of the basic framework of look-up table-based in-loop filtering framework (\textit{LUT-ILF}).} 
	\label{LUT-ILF-P}
	\vspace{-1.5em}
\end{figure*} 

The remainder of this paper is organized as follows. First, we introduce the basic framework (\textit{\textbf{LUT-ILF}}). Second, we introduce the enhanced framework and each module of \textit{\textbf{LUT-ILF-U/V/F}}. Third, we show the comprehensive evaluation of proposed framework. Finally, we discuss the future work of \textit{LUT-ILF} scheme and put forward  future improvements.

\section{Basic Framework of LUT-ILF}
\vspace{-0.1em}
In this section, we introduce the basic framework of \textit{LUT-ILF}. As shown in Fig.~\ref{LUT-ILF-P}, it contains four stages: training filtering network, caching the filtering network into LUT, finetuning of filtering LUT, retrieval of filtering LUT. The cooperation of the above stages realizes the whole filtering process, here we introduce them one by one. 

\textit{\textbf{Stage 1: Training Filtering Network.}} First, due to the size of LUT grows exponentially as the dimension of indexing entries (i.e., target pixel with reference pixels) increases, the lightweight filtering network is trained with the constraint of a small reference range (receptive field, RF) in an end-to-end manner. Here we take the 2$\times$2 reference range (4D LUT) as an example, and the process is shown in stage 1 of Fig.~\ref{LUT-ILF-P}, the \textit{target pixel (to-be-filtered/reconstructed pixel,} $I_0$) \textit{with three surrounding reference pixels (solid line)} serves as the input to the network. To enlarge the size of RF, the rotation ensemble trick is used to cover the 3$\times$3 reference range (dotted line). The final output value (\textit{filtered pixel}) is averaged by all outputs of the 4 rotations ($V_0$$\sim$$V_3$). In training, the filtered and original pixels form a pair, which is supervised by MSE loss.

\textit{\textbf{Stage 2 \& Stage 4: Caching Network into LUT \& Retrieval of Filtering LUT.}} Second, with the network being trained, the 4D LUT is transferred and cached from the output values of the network via traversing all possible inputs (\textit{target pixel with reference pixels}, [$0$$\sim$$255$][$0$$\sim$$255$][$0$$\sim$$255$][$0$$\sim$$255$] for int8 case of input), as shown in stage 2 of Fig.~\ref{LUT-ILF-P}. Note that the storage of LUT with a large input/output range will bring heavy storage cost, for example, the full size of 4D LUT is calculated as $256^4$$\times$1$\times$8 bit $=$ 4096 MB (4 GB), $256^4$ bins for possible input value ($0$$\sim$$255$), 1 for 8-bit output value. To avoid the heavy storage cost, the indexes of full LUT are uniformly sampled and stored in the small LUT (named \textit{Clipped} LUT), which only caches the output value of the \textit{most significant bits} (MSB) of the input pixel value. In our design, the 8-bit input pixel value is uniformly sampled to 4 MSBs, and the 4 MSBs serve as the initial (nearest) index for the indexing of input pixel. The input/output range of indexing is degraded to [$0$,$16$$...$$240$,$255$][$0$,$16$$...$$240$,$255$][$0$,$16$$...$$240$,$255$][$0$,$16$$...$$240$, $255$], and the size of \textit{Clipped} LUT is calculated as $17^4$$\times$1$\times$8 bit $=$ 81.56 KB. In the retrieval process of finetuned filtering LUT, with the indexing of the MSB of input pixels ($I_0,I_1,I_2,I_3$) in 4D \textit{Clipped} LUT, the obtained output values of the nearest index and \textit{least significant bits} (LSB) of the input pixels are used to interpolate the final \textit{retrieved filtered pixel} by linear interpolation model. For the interpolation method of \textit{Clipped} LUT, we follow the same model as \cite{jo2021practical, li2022mulut, liu2023reconstructed, yin2024online, li2024toward}, and use the 4-Simplex interpolation model.

\textit{\textbf{Stage 3: Finetuning of Filtering LUT.}}
To compensate for the degradation of LUT \textit{Clipping}, the finetuning of \textit{Clipped} LUT is performed to adapt to the uniform sampling and the interpolation model, facilitating the interpolation of the final \textit{retrieved filtered pixel} value of non-sampled indexes of LUT from the nearest sampled indexes of LUT. In finetuning, the values of \textit{Clipped} LUT are activated as the trainable parameters and finetuned by the same setting of filtering network training.

\section{LUT-ILF with Reference, Progressive, Weighted Indexing Mechanism}
\begin{figure*}
	\centering
	\vspace{-0.6em}
	\hspace{-9mm}\includegraphics[width=183mm]{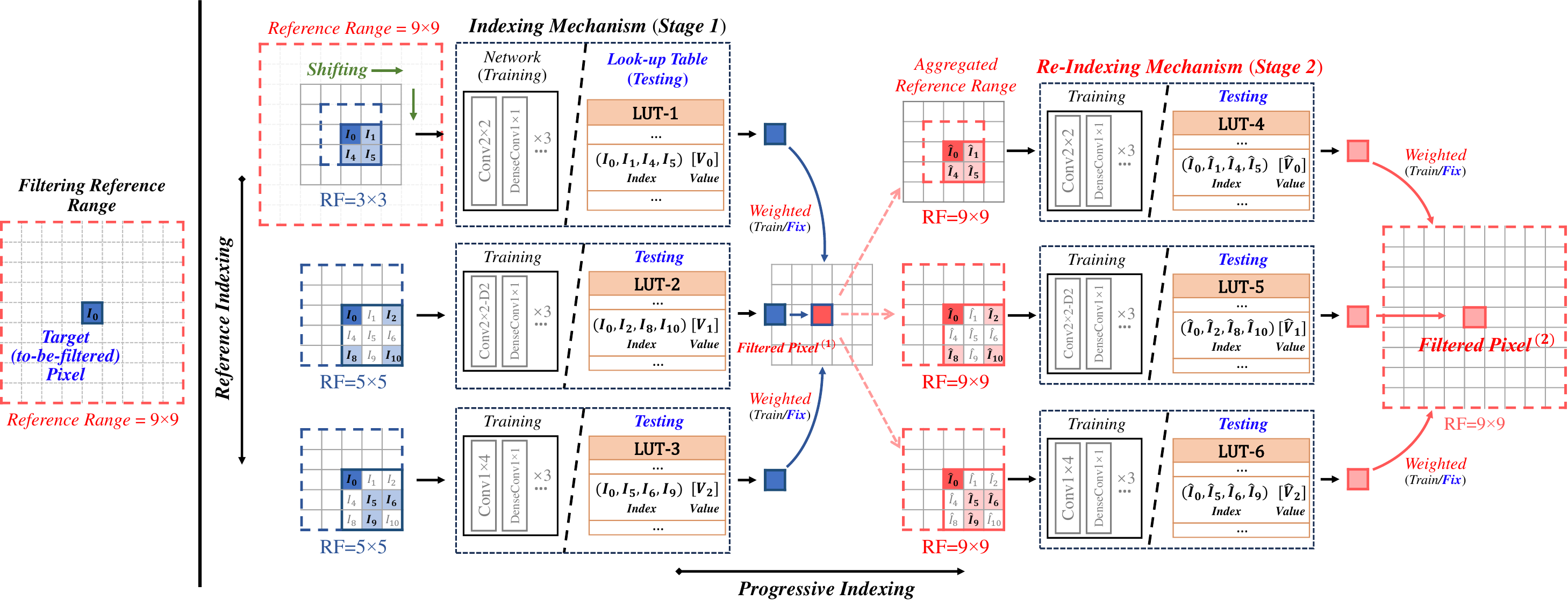}\hspace{-9mm}
	\vspace{-0.8em}
	\caption{Illustration of the \textit{LUT-ILF-V} framework, it contains two parts. On the left, the input (to-be-filtered) pixel with the filtering reference range is shown; On the right, the process of \textit{LUT-ILF-V} is shown, the parallel and cascaded networks/LUTs are performed with \textit{reference} and \textit{progressive indexing} at the training/testing. The covered reference range of each pattern with the rotation trick is marked with dashed boxes. For training, the convolution of each pattern can be implemented with standard convolutions and \textit{unfold/reshape} operations. The Conv2$\times$2-D2 denotes the convolutional layer with a dilation size of 2.}
	\label{co-lut}
	\vspace{-1.5em}
\end{figure*}

\vspace{-0.1em}
For the basic framework (\textit{LUT-ILF}), the efficiency of \textit{LUT-ILF} is mainly subject to two aspects. First, the filtering reference range (RF, only 3$\times$3) is limited with the constraint of LUT size, which is verified as an important factor in traditional filtering tools (such as ALF \cite{tsai2013adaptive} with 7$\times$7 reference range). Second, the selection of reference pixels is very relevant to the filtering (such as the ALF\cite{tsai2013adaptive} with a diamond shape). To address these limitations, inspired by \cite{li2022mulut, li2024toward}, the \textit{\textbf{reference, progressive, weighted indexing mechanism}} is introduced to enhance the above issues. Here, we detail them and serve the \textit{LUT-ILF-V} as an example, the framework is shown in Fig.~\ref{co-lut}.

\begin{figure*}
	\centering
	\vspace{-2.3em}
	\includegraphics[width=162mm]{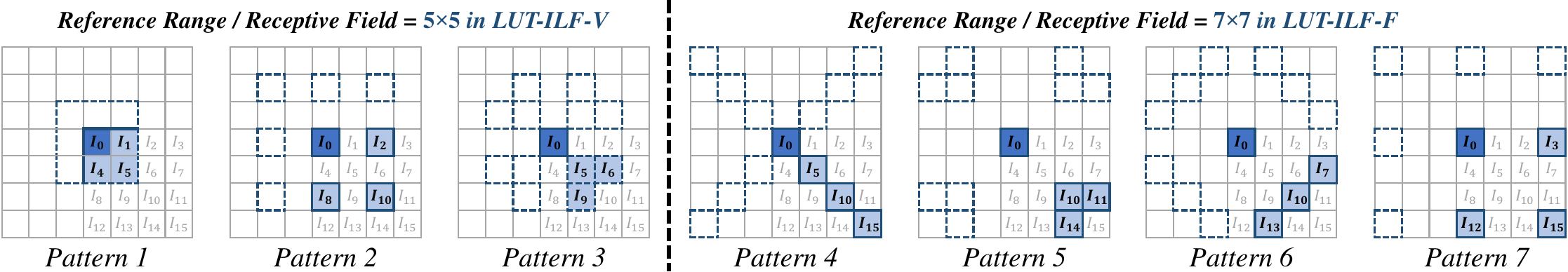}
	\vspace{-0.8em}
	\caption{Illustration of patterns of complementary reference indexing in \textit{LUT-ILF-U} (only \textit{Pattern 1}), \textit{LUT-ILF-V} (\textit{Pattern 1$\sim$3}), and \textit{LUT-ILF-F (\textit{Pattern 1$\sim$7})}. With the use of proposed indexing patterns, \textit{LUT-ILF} can involve and address more reference pixels. For example, with \textit{Pattern 1$\sim$3}, the 5$\times$5 reference range around $I_0$ is fully covered in \textit{LUT-ILF-V}. With \textit{Pattern 1$\sim$7}, the 7$\times$7 reference range around $I_0$ is fully covered in \textit{LUT-ILF-F}. The covered reference pixels with the rotation ensemble trick are marked with dashed boxes.}

	\label{pattern}
	\vspace{-0.5em}
\end{figure*}  

\textit{\textbf{Module 1: Reference Indexing.}} First, the reference pixel range is enlarged to further take advantage of surrounding information for the filtering of target (to-be-filtered) pixel. To avoid the exponential growth in the size of LUT with the dimension, the complementary reference indexing is used to increase the reference range of target pixel by parallelizing more complementary indexing patterns to address more reference pixels and capture the rich local structures. As shown in Fig.~\ref{pattern}, it can cover a wide reference range. In \textit{LUT-ILF-V}, besides the standard indexing pattern of \textit{LUT-ILF} (\textit{Pattern 1}, RF=3$\times$3), complementary \textit{Pattern 2} and \textit{Pattern 3} are used to cover the 5$\times$5 reference range. For the patterns of \textit{LUT-ILF-F}, it can cover the 7$\times$7 reference range. 

In this way, the total size of cached LUTs grows linearly (3 times a 4D \textit{Clipped} LUT, 3$\times$$17^4$$\times$1$\times$8 bit $=$ 244.69 KB), instead of exponentially (the full size of a 25D LUT with an equivalent 5$\times$5 reference range is $256^{(25-4)}$  times a 4D LUT), in a single stage of \textit{reference indexing mechanism}.

\textit{\textbf{Module 2: Progressive Indexing.}} Second, the reference pixel range of the \textit{to-be-filtered pixel} is further enlarged by introducing the cascaded filtering LUTs with \textit{progressive indexing}. As shown in Fig.~\ref{co-lut}, in the whole filtering process of \textit{LUT-ILF-V}, the \textit{re-indexing mechanism} is used to link the cascaded framework between multiple 4D LUTs. In the detailed retrieval process of cascaded filtering LUTs, with the filtering of \textit{target pixel} by multiple indexing patterns (5$\times$5 reference range) in stage 1 of \textit{progressive indexing}, the \textit{filtered pixel} of stage 1 contains the local information of 5$\times$5 reference range implicitly. By \textit{shifting} the filtering window in the 9$\times$9 reference range, the local information of 9$\times$9 reference range can be aggregated into a 5$\times$5 \textit{aggregated reference range}. In stage 2 of \textit{progressive indexing}, the \textit{re-indexing mechanism} can be used to filter the \textit{target pixel} in the \textit{aggregated reference pixels} to achieve the larger reference range implicitly. The process of \textit{progressive indexing} is similar to
cascading multiple convolutional layers in a neural network and achieving information aggregation in the feature domain.

Above these ways, with the utilization of \textit{reference} and \textit{progressive indexing}, the total size of cached LUTs is linear to its indexing capacity (6 times a \textit{Clipped} 4D LUT, 6$\times$$17^4$$\times$1$\times$8 bit $=$ 489.38 KB), instead of exponentially (the full size of an 81D LUT with an equivalent 9$\times$9 reference range is $256^{(81-4)}$ times a 4D LUT), in the whole process of very fast setting of \textit{LUT-ILF} (\textit{LUT-ILF-V}). For the ultrafast setting (\textit{LUT-ILF-U}), the total size of cached LUTs is 0.33$\times$ \textit{LUT-ILF-V}'s size.
For the fast setting (\textit{LUT-ILF-F}), the total size of cached LUTs is 2.3$\times$ \textit{LUT-ILF-V}'s size, instead of exponentially (the full size of a 169D LUT with an equivalent 13$\times$13 reference range is $256^{(169-4)}$ times a 4D LUT). 

\textit{\textbf{Module 3: Learnable Weighting.}} Third, with the extension of reference range, the impact of reference pixels on the \textit{target pixel} should be considered. Instead of the direct average of \textit{filtered pixel} of different indexing patterns, the weights of different indexing patterns are activated as the trained parameters and normalized to [0, 1] with the \textit{softmax}() function to adaptively fit the importance of different reference pixels in the training of filtering network. At the test time, the weights of different patterns are fixed and used by integer operation.

\begin{table*}
	\renewcommand\arraystretch{1.4}
	\centering
	\fontsize{7.3pt}{9pt}\selectfont
	\vspace{-3em}
	\caption{BD-rate and Different Complexity Results of Proposed Method, and Comparison results \\ with the Other In-Loop Filtering Methods under AI and RA Configurations}
	\label{tab:res}
	\vspace{-0.8em}
	\begin{threeparttable}
		\setlength{\tabcolsep}{0.2mm}
		{
\begin{tabular}{c|c|c|c|c|c|c}
	\hline
	\textbf{Methods}                       & \textbf{BD-Rate (AI)} & \textbf{BD-Rate (RA)} & \textbf{Computational Complexity} & \textbf{Storage Cost}        & \textbf{Energy Cost\tnote{2}} & \textbf{Time Complexity (enc/dec, CPU)} \\ \hline
	NNVC-LOP\tnote{1} \, (VTM-11.0)                    & 
	-4.61\%$\sim$-4.78\%                                            & -5.20\%$\sim$-5.37\%  & 17.0 kMACs/pixel                                                                          & \begin{tabular}[c]{@{}c@{}}129.98 KB (\textit{int16})\\ 228.33 KB (\textit{float})\end{tabular} & \begin{tabular}[c]{@{}c@{}}  11900 \textit{pJ} (\textit{int16}) \\ 78200 \textit{pJ} (\textit{float})\end{tabular}                                                                    &\begin{tabular}[c]{@{}c@{}}  108\%/4717\%$\sim$109\%/4724\% (AI) \\ 114\%/8274\%$\sim$114\%/8322\% (RA)\end{tabular}       \\ \hline
	NNVC-HOP\tnote{1} \, (VTM-11.0)                    & -7.79\%$\sim$-7.91\%                                                           &  -10.12\%$\sim$-10.31\%   & 477.0 kMACs/pixel                                                                         & \begin{tabular}[c]{@{}c@{}}2826.2 KB (\textit{int16})\\ 7444.5 KB (\textit{float})\end{tabular}  & \begin{tabular}[c]{@{}c@{}}  333900 \textit{pJ} (\textit{int16}) \\ 2194200 \textit{pJ} (\textit{float})\end{tabular}                                                                   &
	\begin{tabular}[c]{@{}c@{}} 133\%/24372\%$\sim$276\%/134057\% (AI) \\159\%/43509\%$\sim$399\%/227720\% (RA)\end{tabular}     \\ \hline
	\textbf{\textcolor{black}{\textit{LUT-ILF-U} (VTM-11.0)}}   & \textbf{-0.13\%}                                                           &  \textbf{-0.10\%}    & \textbf{0.13 kMACs/pixel}                                                                  & \textbf{164 KB (\textit{int8})\tnote{3}}                                                       & \textbf{180.2 \textit{pJ} }                                                   & \textbf{101\%/102\% (AI), 101\%/105\% (RA)}                                                                  \\ \hline
	\textbf{\textcolor{black}{\textit{LUT-ILF-V} (VTM-11.0)}} & \textbf{-0.34\%}                                                                                                                &   \textbf{-0.27\%}     & \textbf{0.40 kMACs/pixel}                                                                  & \textbf{492 KB (\textit{int8})}                                                       & \textbf{497.2 \textit{pJ}}                                                  & \textbf{102\%/103\% (AI), 103\%/106\% (RA)}                                                                  \\ \hline
	\textbf{\textcolor{black}{\textit{LUT-ILF-F} (VTM-11.0)}} & \textbf{-0.51\%}                                                                                                                 &     \textbf{-0.39\%}     & \textbf{0.93 kMACs/pixel}                                                                  & \textbf{1148 KB (\textit{int8})}                                                      & \textbf{1163.25 \textit{pJ}}                                                 & \textbf{102\%/106\% (AI), 104\%/108\% (RA)}                                                                  \\ \hline
\end{tabular}
		}
	\end{threeparttable}
	\begin{threeparttable}
		\begin{tablenotes}    
			\fontsize{6pt}{8pt}\selectfont
			\item[1] The results of BD-rate, time complexity, computational complexity, storage cost (int/float model) are cited from \cite{AF0043} (LOP) /\cite{AF0014, AF0041} (HOP) and open-sourced repository. 
			\item[2] The energy cost is calculated according to \cite{song2021addersr, sze2017efficient, horowitz20141}. For addition, \textit{int8/int16/float32} corresponds to 0.03/0.05/0.9 $pJ$. For multiplication, the operation of \textit{int8/float16/float32} corresponds to 0.2/1.1/3.7 $pJ$. Since the multiplication of \textit{int16} is not reported in \cite{sze2017efficient}, it is referred to as median of the energy of \textit{int8} and \textit{float16}. For the energy cost of NNVC-ILF, the results are directly calculated by their computational complexity.   
			\item[3] The storage cost of a single model of \textit{LUT-ILF} is shown.
		\end{tablenotes}
	\end{threeparttable}
	\vspace{-2em}
\end{table*}

\textit{\textbf{Summary: General Retrieval Formula.}} Finally, we formulate the retrieval of filtering LUT with utilization of \textit{clipping, reference, progressive, weighted indexing mechanism} in the whole process of \textit{LUT-ILF}. In stage 1 of \textit{LUT-ILF}, for the \textit{target pixel $I_0$ with surrounding reference pixels}, the \textit{filtered pixel} can be addressed and calculated by  
\begin{equation}\label{eq}\fontsize{7.7pt}{6pt}\centering
\begin{aligned}
		Filtered \,\, Pixel^{(1)}  = (W^{(1)}_1 \times LUT^{(1)}_{*p_1}[I_0][I_1][I_4][I_5] + 
		W^{(1)}_2 \times LUT^{(1)}_{*p_2} \\ [I_0][I_2][I_8][I_{10}] 
	    + \text{···} + W^{(1)}_n \times LUT^{(1)}_{*p_n}[\text{·}][\text{·}][\text{·}][\text{·}]\text{···})/n
\end{aligned}
\end{equation}
where $(1)$ denotes the stage number of LUT, $n$ denotes the number of indexing patterns, $LUT_{*}[\text{·}]$ denotes the look-up and interpolation process of LUT retrieval, $Pn$ denotes the pattern ID, $W_n$ denotes the weights of different indexing patterns.

In stage 2, the final \textit{filtered pixel} can be addressed and calculated by
\begin{equation}\label{eq}\fontsize{7.7pt}{6pt}\centering
\begin{aligned}
	Filtered \,\, Pixel^{(2)} = (W^{(2)}_1 \times LUT^{(2)}_{*p_1}[\widehat{I}_0][\widehat{I}_1][\widehat{I}_4][\widehat{I}_5] +
	W^{(2)}_2 \times \\  LUT^{(2)}_{*p_2}[\widehat{I}_0][\widehat{I}_2][\widehat{I}_8][\widehat{I}_{10}] 
	+ \text{···} + W^{(2)}_n \times LUT^{(2)}_{*p_n}[\widehat{\text{·}}][\widehat{\text{·}}][\widehat{\text{·}}][\widehat{\text{·}}]\text{···})/n
\end{aligned}
\end{equation}
where the value ($\widehat{I}$) denotes the output value of the previous filtering stage that serves as the index of the following LUT.

\section{Rate-Distortion-Optimization of LUT-ILF}
\vspace{-0.05em} 
For the integration of \textit{LUT-ILF} into the filtering process of VVC (DBF, SAO, ALF), we set it at the end of all filtering processes, and the decision flag of \textit{LUT-ILF} is signaled in the Coding Tree Unit (CTU) level to indicate the use of proposed method. The flag is determined by the rate-distortion (RD) cost function that $J= SSD +\lambda\times R_{flag}$, where $R_{flag}$ denotes the rates of decision flag in CABAC-based rate estimation, $SSD$ denotes the sum of squared differences (SSD) between the reconstructed result and filtering result of \textit{LUT-ILF}.

\vspace{-0.2em}
\section{Experiment}
\vspace{-0.1em}
In our experiment, the VVC reference software VTM-11.0 is used as the baseline. The codec adopts the configuration of all intra (AI) and random access (RA) according to the VVC Common Test Condition (CTC). The test sequences from classes A to E with different resolutions are tested as specified in \cite{CTCdocument, liu2021jvet}. For each test sequence, quantization parameter (QP) values are set to 22, 27, 32, 37, 42, and Bjontegaard Delta-rate (BD-rate) \cite{2001Calculation} is used as an objective metric to evaluate coding performance. For the complexity metrics, \textit{time complexity}, \textit{computational complexity} (kMACs/pixel \cite{liu2021jvet}), \textit{theoretical energy cost} ($pJ$ \cite{song2021addersr, sze2017efficient, horowitz20141}), and \textit{storage cost} (KB) are evaluated. For the training setup of \textit{LUT-ILF-U/V/F}, as shown in Fig.~\ref{co-lut}\,, the network  is designed as 4 dense convolutions with 1 convolutions in each stage, only the size of the first layer is different to adapt to different shape of pattern, and the BVI-DVC, DIV2K are used as the training datasets\cite{ma2021bvi, agustsson2017ntire}. For different QPs, the LUT is trained separately. The experimental results and the comparison with other methods are shown in Table~\ref{tab:res}. 

\begin{table}
	\renewcommand\arraystretch{1.2}
	\centering
	\fontsize{7.6pt}{9pt}\selectfont
	\caption{Ablation Study of \textit{LUT-ILF-V/F} under AI Configuration}
	\label{tab:ab}
	\vspace{-0.8em}
	\setlength{\tabcolsep}{0.7mm}
	{
		\begin{tabular}{clcccc}
			\hline
			\textbf{Class} &  & \multicolumn{1}{l}{\textbf{\,\,\,\,\,\,\,\,\,\,w/o \textit{PI}}} & \multicolumn{1}{l}{\textbf{\,\,\,\,\,\,\,\,\,\,w/o \textit{LW}}} & \multicolumn{1}{l}{\textbf{\,\,\,\,\,\,\,w/o \textit{RDO}}} & \multicolumn{1}{l}{\textbf{\textit{\,\,\,LUT-ILF-V/F}}} \\ \cline{1-1} \cline{3-6} 
			\textbf{A}              &  &  -0.09\%\,/\,-0.19\%                                & -0.25\%\,/\,-0.39\%                               & 
			1.89\%\,/\,1.06\%                                   & -0.30\%\,/\,-0.45\%                                  \\
			\textbf{B}              &  & -0.08\%\,/\,-0.15\%                                  & -0.19\%\,/\,-0.30\%                                 & 
			2.21\%\,/\,1.32\%                                 & -0.23\%\,/\,-0.40\%                          \\
			\textbf{C}     &  & -0.07\%\,/\,-0.13\%                                  & -0.23\%\,/\,-0.34\%                                  & 
			0.64\%\,/\,0.31\%                                   & -0.29\% / -0.39\%                                   \\
			\textbf{D}     &  & -0.18\%\,/\,-0.27\%                                  & -0.46\%\,/\,-0.60\%                                  & 
			\,0.12\%\,/\,-0.29\%                                   & -0.52\%\,/\,-0.70\%                                   \\
			\textbf{E}     &  & -0.13\%\,/\,-0.21\%                                  & -0.34\%\,/\,-0.59\%                                  & 2.86\%\,/\,1.32\%                  & -0.44\%\,/\,-0.63\%                                   \\ \hline
			\textbf{Avg.}  &  & \textbf{-0.10\%\,/\,-0.17\%}                                   & \textbf{-0.28\%\,/\,-0.43\%}                                  & \textbf{1.55\%\,/\,0.77\%}                                   & \textbf{-0.34\%\,/\,-0.51\%}                                   \\ \hline
		\end{tabular}
	}
	\vspace{-1.6em}
\end{table}

\begin{table}
	\renewcommand\arraystretch{1.25}
	\centering
	\fontsize{7.7pt}{9pt}\selectfont
	\caption{CTU-level Usage Ratio of \textit{LUT-ILF-V/F} under AI Configuration}
	\label{tab:sr}
	\vspace{-0.8em}
	\setlength{\tabcolsep}{1.6mm}
	{
		\begin{tabular}{c|ccccc|c}
			\hline
			\textbf{Class}         & \textbf{A}     & \textbf{B} & \textbf{C} & \textbf{D} & \textbf{E} & \textbf{Avg.} \\ \hline
			\textit{\textbf{LUT-ILF-V}} & 
			31.81\%    & 27.13\%     & 39.44\%         & 49.18\%         & 27.07\%         & \textbf{34.41\%}            \\ \hline
			\textit{\textbf{LUT-ILF-F}} & 43.92\%    & 37.88\%     & 48.10\%         & 58.87\%         & 39.13\%         & \textbf{45.31\%}            \\ \hline
		\end{tabular}
		\vspace{0.9em}
	}

	\renewcommand\arraystretch{1.25}
	\centering
	\fontsize{8.1pt}{9pt}\selectfont
	\caption{BD-rate Results on Low-bitrate Points under AI Configuration}
	\label{tab:lbe}
	\vspace{-0.7em}
	\setlength{\tabcolsep}{1.6mm}
	{
		\begin{tabular}{c|ccccc|c}
			\hline
			\textbf{Class}         & \textbf{A}     & \textbf{B} & \textbf{C} & \textbf{D} & \textbf{E} & \textbf{Avg.} \\ \hline
			\textit{\textbf{LUT-ILF-V}} & -0.70\%    & -0.48\%     & -0.64\%         & -1.08\%         & -0.95\%         & \textbf{-0.74\%}            \\ \hline 
			\textit{\textbf{LUT-ILF-F}} & -1.01\%    & -0.76\%     & -0.84\%         & -1.40\%         & -1.32\%         & \textbf{-1.03\%}            \\ \hline
			
		\end{tabular}
	}
	\vspace{-2em}
\end{table}

\textbf{\textit{Performance Analysis}}: From Table~\ref{tab:res}, we can find that the different modes (\textit{ultrafast, very fast, fast}) of our proposed \textit{LUT-ILF} provide a series of new trade-off points between the performance and efficiency for practical applications. For the quantitative comparisons of performance and complexity, the computational complexity and decoding time complexity of \textit{LUT-ILF} are $130$$\times$$\sim$$3600$$\times$ and $46$$\times$$\sim$$2200$$\times$ lower than  that of popular NN-based ILF methods \cite{AF0043, AF0014, AF0041}, and \textit{LUT-ILF} also shows good performance potential. 

\textbf{\textit{Ablation Study}}: To validate the contributions of core modules in our scheme, we conduct the ablation experiments on proposed \textit{progressive indexing} (\textit{PI}), \textit{learnable weighting} (\textit{LW}), and the CTU-level RDO (\textit{RDO}), under AI configuration. As shown in Table~\ref{tab:ab}, for the comparison of variants and \textit{LUT-ILF}, the results verify the effectiveness of the proposed modules. 

\begin{figure}
	\centering
	\includegraphics[width=75mm]{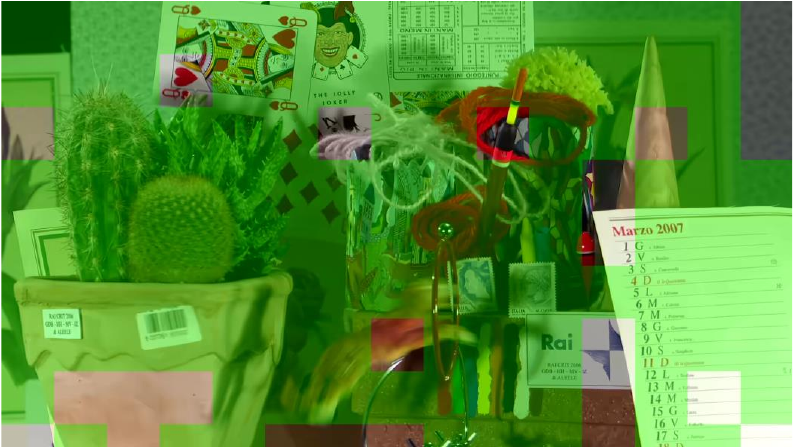}
	\vspace{-0.9em}
	\caption{The selection results of \textit{LUT-ILF-F} of $Cactus\,1920\times1080$ on VTM-11.0 (AI configuration, QP:32, POC:29), the green block indicates the block filtered by \textit{LUT-ILF}.}
	\label{partition}
	\vspace{-1.7em}
\end{figure}

\textbf{\textit{Usage Ratio}}: To verify the efficiency of \textit{LUT-ILF}, we evaluate its usage ratio (Table~\ref{tab:sr}), which is calculated by, $Ratio = {N_{test}}/N_{total}$, where $N_{test}$ indicates the number of filtered CTU, and $N_{total}$ indicates the total number of CTUs. The selection results are also shown in Fig.~\ref{partition}, representing that the \textit{LUT-ILF} can better handle the complex texture regions.

\textbf{\textit{Low-bitrate Points Exploration}}: To further explore the potential of proposed method, we test our proposed method on low bitrate points (QP 27$\sim$47), as shown in Table~\ref{tab:lbe}. The results verify the powerful potential of the proposed method.

\vspace{-0.1em}
\section{Conclusion}
\vspace{-0.1em}
In this paper, we propose an efficient look-up table-based ILF method, which adopts the strong fitting ability of deep neural networks to model the compact look-up tables for ILF. For practical application, the use of \textit{LUT-ILF} does not need to rely on high-performance hardware and devices. The experimental results of \textit{LUT-ILF} demonstrate it can achieve a good performance with low time/computational complexity in VVC, which provides a new practical way for neural network-based video coding tools in the future. For future work, we will further extend the proposed method to improve the performance of more coding tools, such as the interpolation of fractional-pixel motion estimation\cite{yan2018convolutional, yan2019invertibility, li2024object}, reference picture resampling\cite{fu2022efficient}, etc.

\bibliographystyle{IEEEtran}
\bibliography{IEEEexample}
\end{document}


\title{\textit{Supplementary Material}:\\In-Loop Filtering via Trained Look-Up Tables}	

\maketitle

\section{Interpolation Method of Clipped LUT}
\begin{figure}[h]
	\centering
	\includegraphics[width=180mm]{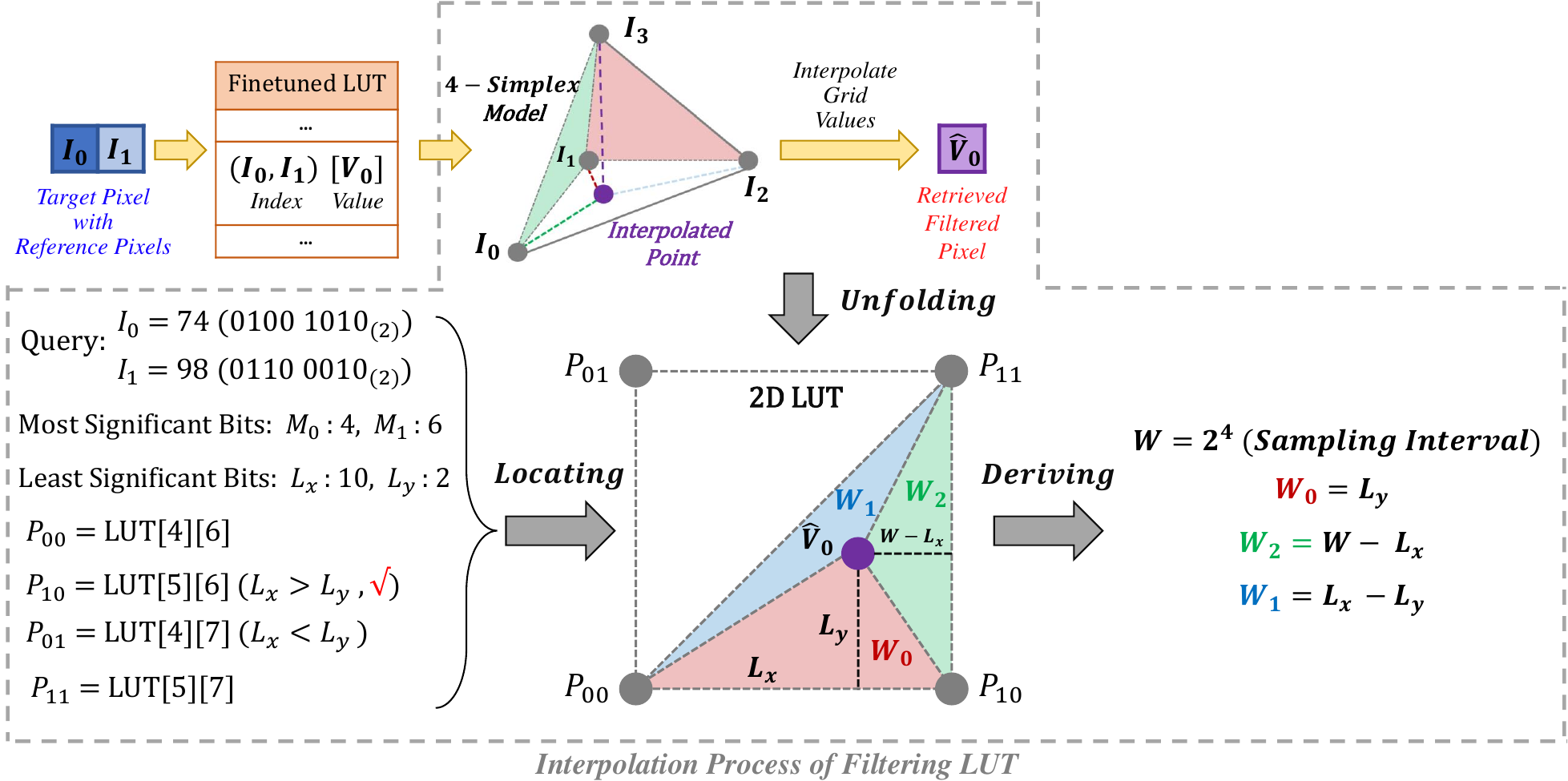}
	\vspace{-0.3em}
	\caption{Illustration of the process of 4-Simplex interpolation method (follow the \cite{jo2021practical}) for \textit{LUT-ILF}.}
	\label{IM}
\end{figure}

To supplement the introduction of the interpolation method of \textit{Clipped} LUT in Section II of the main text, here we detail its process by serving the 2D LUT as an example (Fig.~\ref{IM}\,), which we follow the same model and scheme as \cite{jo2021practical, li2022mulut, liu2023reconstructed, yin2024online, li2024toward}, and use the 4-Simplex interpolation model\cite{kasson1995performing}. Follow the \cite{jo2021practical}, first, for the query value $I_0 = 74\,(0100\,1010_{(2)})$ and $I_1 = 98\,(0110\,0010_{(2)})$ of 2D LUT, as shown in Fig.~\ref{IM}\,, the input values are split into four \textit{most significant
bits} (MSB) and four \textit{least significant bits} (LSB). For the values of MSB, $M_0$ (4) and $M_1$ (6) correspond to $I_0$ and $I_1$, respectively, which is used to determine the initial (nearest) sampled indexes of \textit{Clipped} LUT. For the values of LSB, $L_x$ (10) and $L_y$ (2) correspond to $I_0$ and $I_1$, respectively, which is used to determine the bounding triangle and the weighted factors of bounding vertices. In Fig.~\ref{IM}\,, two bounding vertices are fixed at $P_{00} = LUT[4][6]$ and $P_{11} = LUT[4 + 1][6 + 1]$, and the other vertex is determined by the comparison of $L_x$ and $L_y$. In this case, $P_{10}$ is chosen ($LUT[5][6]$, $L_x > L_y$), otherwise, the $P_{01}$ is chosen. The weight of each vertex is calculated as $w_{0} = L_{y}$, $w_{1} = L_{x} - L_{y}$, $w_{2} = W - L_{x}$, and $W = 2^{4}$ (sampling interval). Finally, the output value is calculated as the weighted sum as follows: $\widehat{V_{0}} = (w_{0}\times P_{11} + w_{1}\times P_{10} + w_{2}\times P_{00})/W$.

Expanding to the 4D case, the 4-Simplex interpolation can be extended to 4D space by using the values of bounding five vertices of 4-simplex geometry \cite{jo2021practical}. The 4-simplex is chosen among a total of 24 cases according to the values of LSB (the cases are detailed in the main text of \cite{jo2021practical} and the open-sourced code of \cite{jo2021practical, li2022mulut}), and the corresponding weights and bounding vertices of each case are used to calculate the final output value.

\section{Detailed Settings and Results of LUT-ILF}

In this section, we supplement the experiment settings and  results of \textit{LUT-ILF}, including the BD-rate results of each sequence/class under all intra (AI) and random access (RA) configurations, the BD-rate results on low-bitrate points, the detailed calculation process of computational complexity and energy cost, and the usage ratio of \textit{LUT-ILF}. 

\subsection{Overall Performance under Common Test Condition on Different Configurations}
\subsubsection{Overall Performance under AI Configuration} Supplementing the Section \uppercase\expandafter{\romannumeral5} of the main text, the R-D performance and usage ratio of the entire \textit{LUT-ILF} framework on the VVC common test sequences is illustrated in Table~\ref{tab:CTC22}\,. Y, U, and V represent the R-D performance gain of the three channels of YUV. We can see that our proposed \textit{LUT-ILF-V/F} can achieve on average, 0.34\% and 0.51\% (marked by bold), and achieve up to 0.76\% and 1.04\% BD-rate reduction (Y component) with high usage ratio and friendly computational complexity (Table I of the main text) on VTM-11.0 for all sequences under AI configuration. The experimental results show that the proposed framework performs better for sequences with abundant texture and complex scenes, such as \textit{DaylightRoad2}, \textit{CatRobot}, \textit{MarketPlace}, and \textit{RaceHorses}. The subjective selection results are mentioned in Section C.

\begin{table}[h]
	\renewcommand\arraystretch{1.5}
	\centering
	\small
	\vspace{-0.5em}
	\caption{BD-rate Results of Our Proposed LUT-ILF Method Compared to VTM-11.0 on CTC Test Sequences \\ with Regular-bitrate Points (QP 22$\sim$42) under All Intra (AI) Configuration.}
	\label{tab:CTC22}
	\vspace{-0.8em}
	\setlength{\tabcolsep}{3.3mm}
	{
		\begin{tabular}{cclcclcc}
			\hline
			\multicolumn{8}{c}{\textbf{All Intra Configuration (\%)}}                                                                                                                                                                                                           \\ \hline
			\multirow{2}{*}{\textbf{Class}}                                                         & \textbf{Sequence}         &                                        & \multicolumn{2}{c}{\textit{\textbf{LUT-ILF-V}}} &  & \multicolumn{2}{c}{\textit{\textbf{LUT-ILF-F}}} \\ \cline{2-2} \cline{4-5} \cline{7-8} 
			& Name                      &                                        & Y                      & \textit{Ratio}         &  & Y                      & \textit{Ratio}         \\ \cline{1-2} \cline{4-5} \cline{7-8} 
			\multirow{4}{*}{\textbf{\begin{tabular}[c]{@{}c@{}}ClassA1\\ (3840x2160)\end{tabular}}} & \textit{Tango2}           &                                        & -0.26\%                & 27.93\%                &  & -0.39\%                & 42.10\%                \\
			& \textit{FoodMarket4}      &                                        & -0.11\%                & 20.81\%                &  & -0.33\%                & 38.25\%                \\
			& \textit{Campfire}         &                                        & -0.30\%                & 33.40\%                &  & -0.37\%                & 44.10\%                \\
			& \textit{\textbf{Average}} &                                        & \textbf{-0.22\%}       & \textbf{27.38\%}       &  & \textbf{-0.36\%}       & \textbf{41.48\%}       \\ \cline{1-2} \cline{4-5} \cline{7-8} 
			\multirow{4}{*}{\textbf{\begin{tabular}[c]{@{}c@{}}ClassA2\\ (3840x2160)\end{tabular}}} & \textit{CatRobot1}        &                                        & -0.47\%                & 37.51\%                &  & -0.67\%                & 48.26\%                \\
			& \textit{DaylightRoad2}    &                                        & -0.38\%                & 35.27\%                &  & -0.50\%                & 45.36\%                \\
			& \textit{ParkRunning3}     &                                        & -0.31\%                & 35.90\%                &  & -0.46\%                & 45.45\%                \\
			& \textit{\textbf{Average}} &                                        & \textbf{-0.39\%}       & \textbf{36.23\%}       &  & \textbf{-0.54\%}       & \textbf{46.35\%}       \\ \cline{1-2} \cline{4-5} \cline{7-8} 
			\multirow{6}{*}{\textbf{\begin{tabular}[c]{@{}c@{}}ClassB\\ (1920x1080)\end{tabular}}}  & \textit{MarketPlace}      &                                        & -0.28\%                & 32.71\%                &  & -0.64\%                & 47.55\%                \\
			& \textit{RitualDance}      &                                        & -0.33\%                & 32.18\%                &  & -0.56\%                & 47.25\%                \\
			& \textit{Cactus}           &                                        & -0.24\%                & 26.31\%                &  & -0.36\%                & 36.98\%                \\
			& \textit{BasketballDrive}  &                                        & -0.02\%                & 12.59\%                &  & -0.09\%                & 21.74\%                \\
			& \textit{BQTerrace}        &                                        & -0.30\%                & 31.84\%                &  & -0.33\%                & 35.89\%                \\
			& \textit{\textbf{Average}} &                                        & \textbf{-0.23\%}       & \textbf{27.13\%}       &  & \textbf{-0.40\%}       & \textbf{37.88\%}       \\ \cline{1-2} \cline{4-5} \cline{7-8} 
			\multirow{5}{*}{\textbf{\begin{tabular}[c]{@{}c@{}}ClassC\\ (832x480)\end{tabular}}}    & \textit{BasketballDrill}  &                                        & -0.33\%                & 37.30\%                &  & -0.42\%                & 42.42\%                \\
			& \textit{BQMall}           & \multicolumn{1}{c}{\textit{}}          & -0.37\%                & 41.02\%                &  & -0.55\%                & 53.85\%                \\
			& \textit{PartyScene}       & \multicolumn{1}{c}{\textit{}}          & -0.22\%                & 40.16\%                &  & -0.30\%                & 48.61\%                \\
			& \textit{RaceHorsesC}      & \multicolumn{1}{c}{\textit{}}          & -0.25\%                & 39.27\%                &  & -0.29\%                & 47.53\%                \\
			& \textit{\textbf{Average}} & \multicolumn{1}{c}{\textit{\textbf{}}} & \textbf{-0.29\%}       & \textbf{39.44\%}       &  & \textbf{-0.39\%}       & \textbf{48.10\%}       \\ \cline{1-2} \cline{4-5} \cline{7-8} 
			\multirow{5}{*}{\textbf{\begin{tabular}[c]{@{}c@{}}ClassD\\ (416x240)\end{tabular}}}    & \textit{BasketballPass}   &                                        & -0.53\%                & 44.91\%                &  & -0.73\%                & 56.58\%                \\
			& \textit{BQSquare}         &                                        & -0.46\%                & 45.08\%                &  & -0.56\%                & 57.08\%                \\
			& \textit{BlowingBubbles}   &                                        & -0.34\%                & 47.50\%                &  & -0.47\%                & 56.91\%                \\
			& \textit{RaceHorses}       &                                        & -0.76\%                & 59.25\%                &  & -1.04\%                & 64.91\%                \\
			& \textit{\textbf{Average}} &                                        & \textbf{-0.52\%}       & \textbf{49.18\%}       &  & \textbf{-0.70\%}       & \textbf{58.87\%}       \\ \cline{1-2} \cline{4-5} \cline{7-8} 
			\multirow{4}{*}{\textbf{\begin{tabular}[c]{@{}c@{}}ClassE\\ (1280x720)\end{tabular}}}   & \textit{FourPeople}       &                                        & -0.51\%                & 36.84\%                &  & -0.66\%                & 51.11\%                \\
			& \textit{Johnny}           & \multicolumn{1}{c}{\textit{}}          & -0.35\%                & 18.74\%                &  & -0.61\%                & 26.71\%                \\
			& \textit{KristenAndSara}   & \multicolumn{1}{c}{\textit{}}          & -0.46\%                & 25.62\%                &  & -0.62\%                & 39.55\%                \\
			& \textit{\textbf{Average}} & \multicolumn{1}{c}{\textit{\textbf{}}} & \textbf{-0.44\%}       & \textbf{27.07\%}       &  & \textbf{-0.63\%}       & \textbf{39.12\%}       \\ \hline
			\multicolumn{2}{c}{\textit{\textbf{Overall}}}                                                                       &                                        & \textbf{-0.34\%}       & \textbf{34.40\%}       &  & \textbf{-0.51\%}       & \textbf{45.30\%}       \\ \hline
		\end{tabular}
	}
\vspace{0.3em}
\end{table}

\subsubsection{Low-bitrate Points Exploration}
To explore the potential of proposed method, we test our proposed method on low bitrate points (QP 27$\sim$47), as shown in Table~\ref{tab:CTC27}\,. Compared to the results of regular QP points (Table~\ref{tab:CTC22}), it further demonstrates the powerful potential and comprehensive effectiveness of the proposed method on the wide range of QP points. Specifically, the proposed \textit{LUT-ILF-V/F} can achieve on average, 0.74\% and 1.03\% (marked by bold), and achieve up to 1.53\% and 1.87\% BD-rate reduction with the high usage ratio (on average 51.92\% and 64.74\%) on VTM-11.0 for all sequences under the AI configuration. 

\begin{table}
	\renewcommand\arraystretch{1.5}
	\centering
	\small
	\vspace{-2em}
	\caption{BD-rate Results of Our Proposed LUT-ILF Method Compared to VTM-11.0 on CTC Test Sequences \\ with Low-bitrate Points (QP 27$\sim$47) under All Intra (AI) Configuration.}
	\label{tab:CTC27}
	\vspace{-0.8em}
	\setlength{\tabcolsep}{3.3mm}
	{
		\begin{tabular}{cclcclcc}
			\hline
			\multicolumn{8}{c}{\textbf{All Intra Configuration (\%)}}                                                                                                                                                                                                                                 \\ \hline
			\multirow{2}{*}{\textbf{Class}}                                                         & \textbf{Sequence}         &                                        & \multicolumn{2}{c}{\textit{\textbf{LUT-ILF-V}}}         &  & \multicolumn{2}{c}{\textit{\textbf{LUT-ILF-F}}} \\ \cline{2-2} \cline{4-5} \cline{7-8} 
			& Name                      &                                        & Y                & \textit{Ratio}                       &  & Y                      & \textit{Ratio}         \\ \cline{1-2} \cline{4-5} \cline{7-8} 
			\multirow{4}{*}{\textbf{\begin{tabular}[c]{@{}c@{}}ClassA1\\ (3840x2160)\end{tabular}}} & \textit{Tango2}           &                                        & -0.61\%          & 47.05\%                              &  & -0.95\%                & 61.72\%                \\
			& \textit{FoodMarket4}      &                                        & -0.21\%          & 40.35\%                              &  & -0.55\%                & 57.97\%                \\
			& \textit{Campfire}         &                                        & -0.64\%          & 44.45\%                              &  & -0.84\%                & 63.07\%                \\
			& \textit{\textbf{Average}} &                                        & \textbf{-0.49\%} & \textbf{43.95\%}                     &  & \textbf{-0.78\%}       & \textbf{60.92\%}       \\ \cline{1-2} \cline{4-5} \cline{7-8} 
			\multirow{4}{*}{\textbf{\begin{tabular}[c]{@{}c@{}}ClassA2\\ (3840x2160)\end{tabular}}} & \textit{CatRobot1}        &                                        & -0.98\%          & 54.26\%                              &  & -1.40\%                & 67.08\%                \\
			& \textit{DaylightRoad2}    &                                        & -0.79\%          & 54.50\%                              &  & -1.12\%                & 64.96\%                \\
			& \textit{ParkRunning3}     &                                        & -0.96\%          & 54.75\%                              &  & -1.20\%                & 65.41\%                \\
			& \textit{\textbf{Average}} &                                        & \textbf{-0.91\%} & \textbf{54.50\%}                     &  & \textbf{-1.24\%}       & \textbf{65.82\%}       \\ \cline{1-2} \cline{4-5} \cline{7-8} 
			\multirow{6}{*}{\textbf{\begin{tabular}[c]{@{}c@{}}ClassB\\ (1920x1080)\end{tabular}}}  & \textit{MarketPlace}      &                                        & -0.68\%          & 51.74\%                              &  & -1.08\%                & 67.37\%                \\
			& \textit{RitualDance}      &                                        & -0.49\%          & 49.10\%                              &  & -1.02\%                & 65.60\%                \\
			& \textit{Cactus}           &                                        & -0.49\%          & 41.77\%                              &  & -0.73\%                & 55.19\%                \\
			& \textit{BasketballDrive}  &                                        & -0.10\%          & 25.65\%                              &  & -0.24\%                & 38.51\%                \\
			& \textit{BQTerrace}        &                                        & -0.66\%          & 44.98\%                              &  & -0.76\%                & 50.63\%                \\
			& \textit{\textbf{Average}} &                                        & \textbf{-0.48\%} & \textbf{42.65\%}                     &  & \textbf{-0.76\%}       & \textbf{55.46\%}       \\ \cline{1-2} \cline{4-5} \cline{7-8} 
			\multirow{5}{*}{\textbf{\begin{tabular}[c]{@{}c@{}}ClassC\\ (832x480)\end{tabular}}}    & \textit{BasketballDrill}  &                                        & -0.77\%          & 57.23\%                              &  & -1.00\%                & 62.40\%                \\
			& \textit{BQMall}           & \multicolumn{1}{c}{\textit{}}          & -0.70\%          & 60.61\%                              &  & -1.02\%                & 73.47\%                \\
			& \textit{PartyScene}       & \multicolumn{1}{c}{\textit{}}          & -0.53\%          & 59.64\%                              &  & -0.71\%                & 68.09\%                \\
			& \textit{RaceHorsesC}      & \multicolumn{1}{c}{\textit{}}          & -0.55\%          & 47.73\%                              &  & -0.64\%                & 69.88\%                \\
			& \textit{\textbf{Average}} & \multicolumn{1}{c}{\textit{\textbf{}}} & \textbf{-0.64\%} & \textbf{56.30\%}                     &  & \textbf{-0.84\%}       & \textbf{68.46\%}       \\ \cline{1-2} \cline{4-5} \cline{7-8} 
			\multirow{5}{*}{\textbf{\begin{tabular}[c]{@{}c@{}}ClassD\\ (416x240)\end{tabular}}}    & \textit{BasketballPass}   &                                        & -0.97\%          & 64.91\%                              &  & -1.37\%                & 76.58\%                \\
			& \textit{BQSquare}         &                                        & -1.11\%          & 65.08\%                              &  & -1.41\%                & 77.08\%                \\
			& \textit{BlowingBubbles}   &                                        & -0.69\%          & 67.50\%                              &  & -0.93\%                & 76.91\%                \\
			& \textit{RaceHorses}       &                                        & -1.53\%          & 79.25\%                              &  & -1.87\%                & 84.91\%                \\
			& \textit{\textbf{Average}} &                                        & \textbf{-1.08\%} & \textbf{69.18\%}                     &  & \textbf{-1.40\%}       & \textbf{78.87\%}       \\ \cline{1-2} \cline{4-5} \cline{7-8} 
			\multirow{4}{*}{\textbf{\begin{tabular}[c]{@{}c@{}}ClassE\\ (1280x720)\end{tabular}}}   & \textit{FourPeople}       &                                        & -1.04\%          & 56.50\%                              &  & -1.47\%                & 71.10\%                \\
			& \textit{Johnny}           & \multicolumn{1}{c}{\textit{}}          & -0.73\%          & 33.14\%                              &  & -1.12\%                & 46.24\%                \\
			& \textit{KristenAndSara}   & \multicolumn{1}{c}{\textit{}}          & -1.08\%          & 45.24\%                              &  & -1.36\%                & 59.49\%                \\
			& \textit{\textbf{Average}} & \multicolumn{1}{c}{\textit{\textbf{}}} & \textbf{-0.95\%} & \textbf{44.96\%}                     &  & \textbf{-1.32\%}       & \textbf{58.94\%}       \\ \hline
			\multicolumn{2}{c}{\textit{\textbf{Overall}}}                                                                       &                                        & \textbf{-0.74\%} & \multicolumn{1}{l}{\textbf{51.92\%}} &  & \textbf{-1.03\%}       & \textbf{64.74\%}       \\ \hline
		\end{tabular}
	}
\vspace{-1em}
\end{table}

\subsubsection{Inter-Frame Configuration Exploration}
To further verify the \textbf{\textit{in-loop capacity}} (reference dependent capability) of our proposed method, here we supplement the detailed BD-rate result of the proposed method (\textit{LUT-ILF-V/F}) under random access (RA) configuration, as shown in Table~\ref{tab:RAA}\,. From the quantitative comparisons of performance and complexity under inter-frame configuration (Table~\ref{tab:RAA} and Table I of the main text), the \textit{LUT-ILF} also shows good performance potential with friendly computational complexity and energy cost, which further verifies the better in-loop capacity of the proposed \textit{LUT-ILF} framework. 

\begin{table}[h]
	\renewcommand\arraystretch{1.4}
	\centering
	\small
	\vspace{-2em}
	\caption{BD-rate Results of Our Proposed LUT-ILF Method Compared to VTM-11.0 on CTC Test Sequences \\ with Regular-bitrate Points (QP 22$\sim$42) under Random Access (RA) Configuration.}
	\label{tab:RAA}
	\vspace{-0.8em}
	\setlength{\tabcolsep}{3.3mm}
	{
\begin{tabular}{cclclc}
	\hline
	\multicolumn{6}{c}{\textbf{Random Access Configuration (\%)}}                                                                                                                         \\ \hline
	\multirow{2}{*}{\textbf{Class}}                                                         & \textbf{Sequence}         &  & \textit{\textbf{LUT-ILF-V}} &  & \textit{\textbf{LUT-ILF-F}} \\ \cline{2-2} \cline{4-4} \cline{6-6} 
	& Name                      &  & Y                           &  & Y                           \\ \cline{1-2} \cline{4-4} \cline{6-6} 
	\multirow{4}{*}{\textbf{\begin{tabular}[c]{@{}c@{}}ClassA1\\ (3840x2160)\end{tabular}}} & \textit{Tango2}           &  & -0.27\%                           &  & -0.39\%                           \\
	& \textit{FoodMarket4}      &  & -0.13\%                            &  & -0.10\%                           \\
	& \textit{Campfire}         &  & -0.20\%                            &  & -0.14\%                           \\
	& \textit{\textbf{Average}} &  & \textbf{-0.20\%}                  &  & \textbf{-0.21\%}                  \\ \cline{1-2} \cline{4-4} \cline{6-6} 
	\multirow{4}{*}{\textbf{\begin{tabular}[c]{@{}c@{}}ClassA2\\ (3840x2160)\end{tabular}}} & \textit{CatRobot1}        &  & -0.30\%                           &  & -0.59\%                           \\
	& \textit{DaylightRoad2}    &  & -0.25\%                           &  & -0.53\%                           \\
	& \textit{ParkRunning3}     &  & -0.21\%                           &  & -0.23\%                           \\
	& \textit{\textbf{Average}} &  & \textbf{-0.25\%}                  &  & \textbf{-0.45\%}                  \\ \cline{1-2} \cline{4-4} \cline{6-6} 
	\multirow{6}{*}{\textbf{\begin{tabular}[c]{@{}c@{}}ClassB\\ (1920x1080)\end{tabular}}}  & \textit{MarketPlace}      &  & -0.26\%                           &  & -0.66\%                            \\
	& \textit{RitualDance}      &  & -0.27\%                           &  & -0.20\%                            \\
	& \textit{Cactus}           &  & -0.25\%                           &  & -0.37\%                            \\
	& \textit{BasketballDrive}  &  & -0.10\%                           &  & -0.19\%                            \\
	& \textit{BQTerrace}        &  & -0.19\%                           &  & -0.15\%                            \\
	& \textit{\textbf{Average}} &  & \textbf{-0.21\%}                  &  & \textbf{-0.31\%}                  \\ \cline{1-2} \cline{4-4} \cline{6-6} 
	\multirow{5}{*}{\textbf{\begin{tabular}[c]{@{}c@{}}ClassC\\ (832x480)\end{tabular}}}    & \textit{BasketballDrill}  &  & -0.23\%                           &  & -0.25\%                            \\
	& \textit{BQMall}           &  & -0.13\%                           &  & -0.25\%                            \\
	& \textit{PartyScene}       &  & -0.09\%                           &  & -0.17\%                            \\
	& \textit{RaceHorsesC}      &  & -0.13\%                           &  & -0.09\%                            \\
	& \textit{\textbf{Average}} &  & \textbf{-0.14\%}                  &  & \textbf{-0.19\%}                  \\ \cline{1-2} \cline{4-4} \cline{6-6} 
	\multirow{5}{*}{\textbf{\begin{tabular}[c]{@{}c@{}}ClassD\\ (416x240)\end{tabular}}}    & \textit{BasketballPass}   &  & -0.38\%                           &  & -0.44\%                            \\
	& \textit{BQSquare}         &  & -0.48\%                           &  & -0.61\%                            \\
	& \textit{BlowingBubbles}   &  & -0.22\%                           &  & -0.38\%                            \\
	& \textit{RaceHorses}       &  & -0.30\%                           &  & -0.55\%                            \\
	& \textit{\textbf{Average}} &  & \textbf{-0.35\%}                  &  & \textbf{-0.49\%}                  \\ \cline{1-2} \cline{4-4} \cline{6-6} 
	\multirow{4}{*}{\textbf{\begin{tabular}[c]{@{}c@{}}ClassE\\ (1280x720)\end{tabular}}}   & \textit{FourPeople}       &  & -0.61\%                           &  & -0.80\%                            \\
	& \textit{Johnny}           &  & -0.35\%                           &  & -0.63\%                            \\
	& \textit{KristenAndSara}   &  & -0.41\%                           &  & -0.73\%                            \\
	& \textit{\textbf{Average}} &  & \textbf{-0.46\%}                  &  & \textbf{-0.72\%}                  \\ \hline
	\multicolumn{2}{c}{\textit{\textbf{Overall}}}                                                                       &  & \textbf{-0.27\%}                  &  & \textbf{-0.39\%}                  \\ \hline
\end{tabular}
	}
\end{table}

\subsection{Computational Complexity and Theoretical Energy Cost of LUT-ILF}

To clearly show the calculation process of the computational complexity  of the proposed method and facilitate researchers to follow, here we detail the specific operation num of the basic framework (\textit{LUT-ILF-U/V}) on the pixel (per pixel) and frame level (a 1920 × 1080 HD frame), as shown in Table~\ref{tab:cc}\,. For the extension of the proposed framework, computational complexity can be calculated with the reference of these two basic schemes for relative expansion, such as the proposed \textit{LUT-ILF-F}.

For the reported computational complexity results in Table I of the main text, it's worth noting that the reported results indicate the “worse-case” computational complexity, which means that if the number of additions is more than the number of multiplications, the calculation of kMAC tends to be additions. In practice, the computational complexity and energy cost of the proposed method are \textbf{\textit{lower}} than the reported results, because the multiplication cost is considered much more than the addition for practical application\cite{song2021addersr, sze2017efficient, horowitz20141}.

\begin{table*}
	\renewcommand\arraystretch{1.8}
	\centering
	\small
	\caption{The Computational Complexity Results and Specific Operation Num of Our Proposed Basic Framework \\  (\textit{LUT-ILF-U/V}) on the Pixel (per pixel) and Frame level (a 1920 × 1080 HD frame).}
	\label{tab:cc}
	\vspace{-0.8em}
	\setlength{\tabcolsep}{3.3mm}
	\begin{threeparttable}
	\setlength{\tabcolsep}{2.2mm}
	{
	\begin{tabular}{cccc}
	\hline
	\multicolumn{4}{c}{\textbf{Detailed Calculation of Computational Complexity / Energy Cost of LUT-ILF-U/V/F}}                                                                                                                                                           \\ \hline
	\multicolumn{1}{c|}{\textbf{Operation}}                                                                                   & \multicolumn{1}{c|}{\textbf{Level}}                       & \multicolumn{1}{c|}{\textit{\textbf{Operation Num of LUT-ILF-U}}} & \textit{\textbf{Operation Num of LUT-ILF-V}} \\ \hline
	\multicolumn{1}{c|}{int8 Add}                                                                                             & \multicolumn{1}{c|}{\multirow{6}{*}{Pixel-wise}}          & \multicolumn{1}{c|}{70}                          & 206                         \\ \cline{1-1}
	\multicolumn{1}{c|}{int8 Multiply}                                                                                        & \multicolumn{1}{c|}{}                                     & \multicolumn{1}{c|}{4}                           & 4                           \\ \cline{1-1}
	\multicolumn{1}{c|}{int32 Add}                                                                                            & \multicolumn{1}{c|}{}                                     & \multicolumn{1}{c|}{68}                          & 190                         \\ \cline{1-1}
	\multicolumn{1}{c|}{int32 Multiply}                                                                                       & \multicolumn{1}{c|}{}                                     & \multicolumn{1}{c|}{55}                          & 152                         \\ \cline{1-1}
	\multicolumn{1}{c|}{Total Add}                                                                                            & \multicolumn{1}{c|}{}                                     & \multicolumn{1}{c|}{138}                         & 396                         \\ \cline{1-1}
	\multicolumn{1}{c|}{Total Multiply}                                                                                       & \multicolumn{1}{c|}{}                                     & \multicolumn{1}{c|}{59}                          & 156                         \\ \hline
	\multicolumn{1}{c|}{int8 Add}                                                                                             & \multicolumn{1}{c|}{\multirow{4}{*}{Frame-wise}}          & \multicolumn{1}{c|}{145,152,000}                 & 427,161,600                 \\ \cline{1-1}
	\multicolumn{1}{c|}{int8 Multiply}                                                                                        & \multicolumn{1}{c|}{}                                     & \multicolumn{1}{c|}{8,294,400}                   & 8,294,400                   \\ \cline{1-1}
	\multicolumn{1}{c|}{int32 Add}                                                                                            & \multicolumn{1}{c|}{}                                     & \multicolumn{1}{c|}{141,004,800}                 & 393,984,000                 \\ \cline{1-1}
	\multicolumn{1}{c|}{int32 Multiply}                                                                                       & \multicolumn{1}{c|}{}                                     & \multicolumn{1}{c|}{114,048,000}                 & 315,187,200                 \\ \hline
	\multicolumn{1}{c|}{\textbf{Total Add}}                                                                                   & \multicolumn{1}{c|}{\multirow{2}{*}{\textbf{Frame-wise}}} & \multicolumn{1}{c|}{\textbf{286,156,800}}        & \textbf{821,145,600}        \\ \cline{1-1} \cline{3-4} 
	\multicolumn{1}{c|}{\textbf{Total Multiply}}                                                                              & \multicolumn{1}{c|}{}                                     & \multicolumn{1}{c|}{\textbf{122,342,400}}        & \textbf{323,481,600}        \\ \hline
	\multicolumn{1}{c|}{\textbf{\begin{tabular}[c]{@{}c@{}}Worse-case Computational\\ Complexity\tnote{1} \,(kMACs/pixel)\end{tabular}}} & \multicolumn{1}{c|}{\textbf{Pixel-wise}}                  & \multicolumn{1}{c|}{\textbf{0.13}}               & \textbf{0.40}               \\ \hline
	\multicolumn{1}{c|}{\textbf{Energy Cost\tnote{2}  \,(pJ)}}                                                                            & \multicolumn{1}{c|}{\textbf{Frame-wise}}                  & \multicolumn{1}{c|}{\textbf{180.2}}              & \textbf{497.2}              \\ \hline
	\end{tabular}
	}
\end{threeparttable}
\begin{threeparttable}
\begin{tablenotes}    
\fontsize{7pt}{8pt}\selectfont
\item[1] The “worse-case” means that if the number of
additions is more than the number of multiplications, the calculation of kMAC tends to be additions.
\item[2] The energy cost is calculated according to \cite{song2021addersr, sze2017efficient, horowitz20141}. For addition, \textit{int8/int16/float32} corresponds to 0.03/0.05/0.9 $pJ$. For multiplication, the operation of \textit{int8/float16/float32} corresponds to 0.2/1.1/3.7 $pJ$.  
\end{tablenotes}
\end{threeparttable}
\end{table*}

\subsection{Usage Ratio of LUT-ILF}

\begin{figure}[h]   	
	\centering
	\small
	\includegraphics[width=170mm]{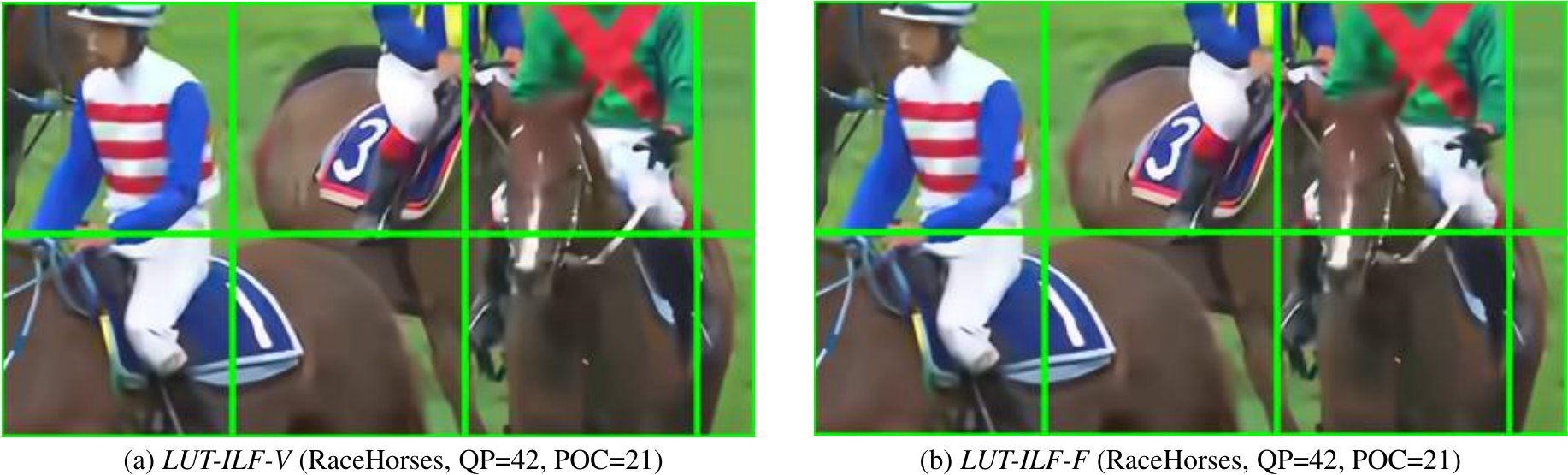}\vspace{1em}
	
	\includegraphics[width=170mm]{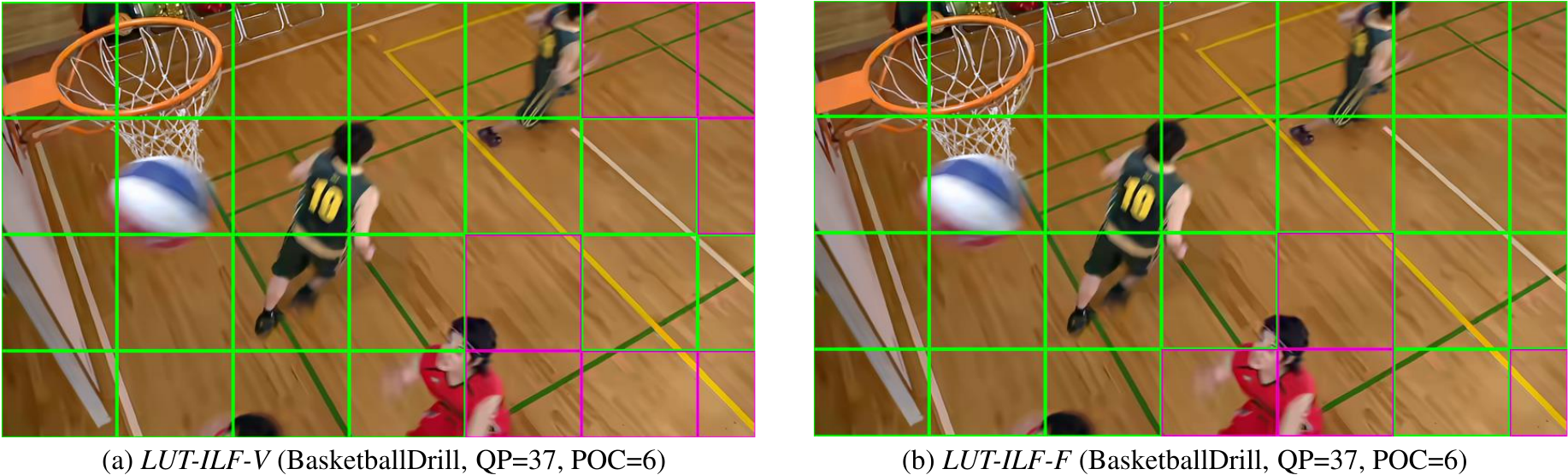}
	\vspace{1em}
	
	\includegraphics[width=170mm]{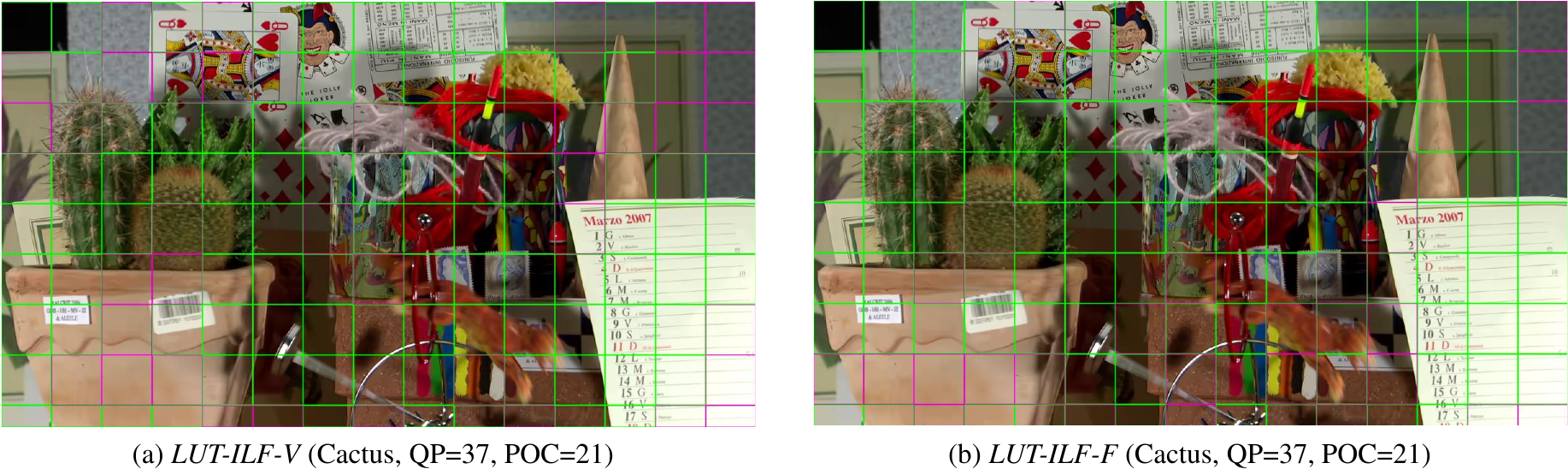}
	
	\caption{The usage results of \textit{LUT-ILF-V/F} of CTC test sequences on VTM-11.0 (Configuration: AI), where (a) uses the \textit{LUT-ILF-V} filter, (b) uses the \textit{LUT-ILF-F} filter. The green block indicates the block filtered by \textit{LUT-ILF}, and the purple block indicates the opposite.}
	\label{fig:srp}
\end{figure}

In Fig.~\ref{fig:srp}\,, here we show the CTU-level usage results of the proposed \textit{LUT-ILF-V/F} to exhibit the usage scenarios. The green block indicates the block filtered by \textit{LUT-ILF} , and the purple block indicates the opposite.

\bibliographystyle{IEEEtran}
\bibliography{IEEEexample-sup}